# Simulation Study of TenTen: A new Multi-TeV IACT array

V. STAMATESCU[1], G. ROWELL[1], R. CLAY[1], B. DAWSON[1], J. DENMAN[1], R. DUNBAR[1], R. PROTHEROE[1], A. SMITH[1], G. THORNTON[1], N. WILD[1]

[1]*University of Adelaide*
*victor.stamatescu@adelaide.edu.au*

**Abstract:** TenTen is a proposed array of Imaging Atmospheric Cherenkov Telescopes (IACT) optimized for the gamma ray energy regime of 10 TeV to 100 TeV, but with a threshold of ~1 to a few TeV. It will offer a collecting area of 10 km$^2$ above energies of 10 TeV. In the initial phase, a cell of 3 to 5 modest-sized telescopes, each with 10-30 m$^2$ mirror area, is suggested for an Australian site. A possible expansion of the array could comprise many such cells. Here we present work on configuration and technical issues from our simulation studies of the array. Working topics include array layout, telescope size and optics, camera field of view, telescope trigger system, electronics, and site surveys.

## Introduction

We present design work for an IACT telescope array operating in the energy regime of 1 TeV to 100 TeV, known as TenTen, the motivation for which is studied in a companion paper [13]. Conducting multi-TeV astronomy within reasonable observation times will require a large effective collection area of ~10 km$^2$ at energies above 10 TeV ([13] and references therein). To this end, we investigate the performance of an array of many modest-sized telescopes employing established ideas from existing stereoscopic instruments such as H.E.S.S. [7], VERITAS [18], CANGAROO III [2] and MAGIC II [9].

Following Plyasheshnikov et al. [10], initial simulations have been carried out [12] by adopting a HEGRA-type cell approach [11]. While this may not be optimum, it was considered appropriate for first-order studies. The basic parameters of the cell such as telescope spacing, camera field of view (FoV) and mirror optics were tuned to improve the effective collection area.

A camera electronics design based on existing technology is also under consideration.

## Initial Simulation Study

Simulations of extensive air showers were performed with CORSIKA v6.204 [6] and SYBILL [5]. On-axis gamma ray-induced air showers with a zenith angle of 30°, as well as isotropic proton-induced air showers, between 1 TeV and 100 TeV, were simulated using flat primary energy spectra.

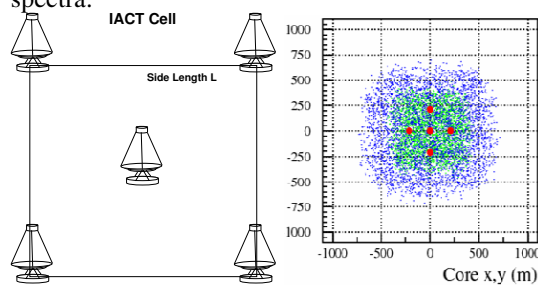

Figure 1: Left: Layout of an IACT cell with side length L. Right: True core locations of triggered showers for L = 300 m, see text for trigger details. Large red dots represent telescope positions, green points represent 1-10 TeV events, blue points represent 10-100 TeV events.

An example cell layout with 5 IACT telescopes was simulated (Figure 1, left), with four tele-scopes placed at the corners of a square of side L, and the fifth at its centre. The observation level was 200 m a.s.l., which is appropriate for Australian sites. Given our higher energy threshold, a near sea level site is favorable as it leads to a larger collecting area for energies above 10 TeV. The telescope response was simulated using [15] and verified with an independent telescope simulation based on [3].

Simulated telescopes used f/1.5 optics and a 23.8 m$^2$ mirror comprising 84 x 60 cm spherical mirror facets. An elliptical dish profile was simulated as obtained in [14], using conic parameters $\delta = 5$ and $r = 0.85f$, due to its superior off-axis performance in comparison to a Davies Cotton design. In addition, mirror segments were canted so that on-axis light was imaged at the focus, which gave a reduced optical point spread function by a factor of a few for small off-axis angles. The focal ratio used allowed the containment of at least 80% of photons from a point source inside a pixel of 0.25$^\circ$ diameter, for light arriving up to 4$^\circ$ off-axis (Figure 2). For comparison, the d80 values obtained using a Davies Cotton design were ~10% larger for off axis angles in the 4$^\circ$ to 5$^\circ$ range.

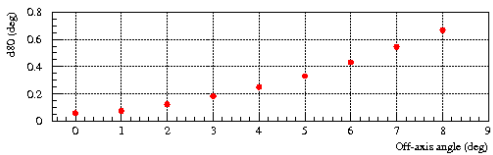

Figure 2: Parameter d80 (deg.), the 80% containment diameter, versus off-axis angle (deg.).

A square camera made of 32x32 square pixels with side 0.25$^\circ$ was modelled with the above mentioned optics. The resulting FoV was 8.2$^\circ$x8.2$^\circ$, including dead space. The camera response was simulated using conservative values for the camera trigger (2 next-neighbor pixels above 12 p.e.) and a small 16 ns ADC gate centered on the camera trigger time (similar to the H.E.S.S. electronics). The hardware simulation required that at least two telescopes met the camera trigger condition. Stereoscopic reconstruction of the showers was performed if more than two images had *size* > 60 p.e. and *dist* < 3.5$^\circ$ (distance of the centre of gravity from the FoV centre).

With L set to 300 m, a single cell was found to have an energy threshold of ~2 TeV. The effective collection area of the simulated cell is given in Figure 3 and shows >1 km$^2$ attained for >30 TeV. Also represented is the H.E.S.S. effective collection area for 30$^\circ$ zenith angle, after 'std cuts' [1] and removal of mean scaled width (MSW) [8] cut efficiencies. The effective collection area of the cell can be estimated to be larger than that of H.E.S.S. by a factor of ~2 at 10 TeV and by a factor of ~5 at 100 TeV. Extrapolating to an array of 10 independent cells (such that there are no common events between cells), we estimate the total effective collection area to be ~10 km$^2$.

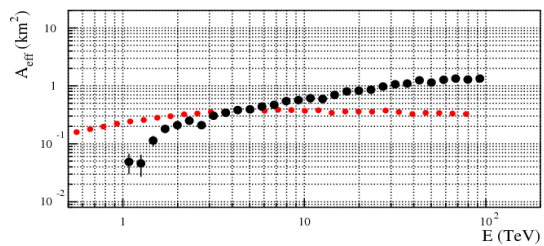

Figure 3: Effective collection area for an IACT cell with L = 300 m at 200 m a.s.l., (larger points) and the equivalent (see text) H.E.S.S. effective collection area (smaller points)

Performance parameters of a cell for energies above 10 TeV, such as cosmic ray background rejection (based on MSW), angular resolution (a few arc minutes) and energy resolution were found to be comparable to those of HEGRA [11] and H.E.S.S. [1] operating in their respective energy ranges. While these simulation results confirm that our initial approach works well, we are currently refining the telescope simulation for detection above 10 TeV.

## Refinement of simulations

The large FoV in our simulated cell permits the detection of showers out to large core distances (> 200 m) from the telescopes (Figure 1, right). As core distances become larger, the shower images are more elongated and take a longer time to form in the camera. For on-axis showers with core distance of 100 m, Cherenkov light arrives within ~14 ns, whereas for showers at 400 m the image can take ~100 ns to form in the focal plane (Figure 4).

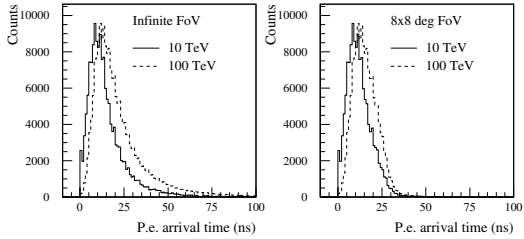

Figure 4: Arrival time distributions w.r.t. the first p.e. to reach the telescope, for two different energies, and for an infinite FoV (left) and an 8x8° FoV (right). Showers had a fixed core distance of 400 m, a zenith angle $30^0$ and arrived on-axis.

In the right panel of Figure 4, the finite FoV limits the maximum time spread of an image for on-axis showers because the 'tails' of images are truncated. However, larger time spreads, like those in the left panel, can be expected when the image is fully contained in the FoV (e.g. for off axis gamma rays). Detected photons with later arrival times (>30 ns after the first p.e.) were found to be more scattered in the focal plane and mainly situated in the 'tail' of the image.

Given the maximum time spread of the images, an integration gate in each pixel that centered on a camera-wide trigger would need to be order of 100 ns to contain the Cherenkov signal. Such large gate widths would result in increased night sky background contamination (NSB) of the data. This problem may be overcome with a 'time-floating' ADC gate. Given that the mean of the distribution of arrival time RMS in any pixel was found to be fairly independent of core distance and only increased slightly with an order of magnitude increase in primary energy, the use of a small 10 to 20 ns gate triggered off its own PMT pulse seems ideal.

The following block diagrams (Figures 5 and 6) represent two possible readout electronics concepts for a single channel, as well as a possible camera trigger. This trigger takes the short pulses from all channels and makes the decision whether to read out the data based on 2 levels of discrimination: the first is a threshold on the number of p.e. that is applied to each pixel, the second is the requirements of next-neighbor (NN) multiplicity. In both cases the trigger signals are delayed to allow sufficient time for the image to form in the camera.

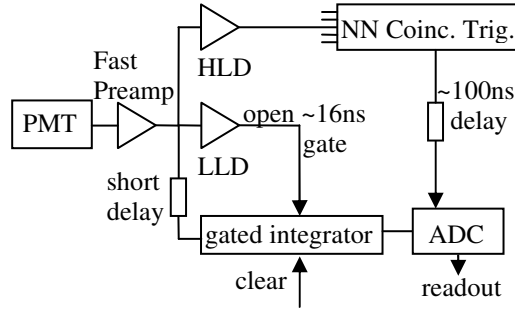

Figure 5: Block diagram representing the first option as described in the text.

Concerning the signal digitization and readout, the first option (Figure 5) uses two discriminators per pixel, one with a higher threshold (HLD) used to build the camera trigger and a lower level discriminator (LLD) which opens a small time gate over which the PMT pulse is integrated. The analogue value is held until a resultant camera trigger arrives, at which time it is digitized and read out and the analogue integral value is cleared. If the LLD triggered but no camera trigger arrives within ~200 ns, the analogue integral value is cleared. To prevent single pixel triggers due to NSB pileup becoming part of an image if the shower arrives within ~200 ns, it may be necessary to include a second gated integrator (not shown) that is digitized instead if the HLD is triggered.

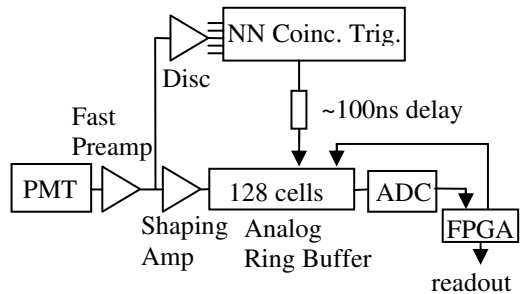

Figure 6: Block diagram representing the second option as described in the text.

In the second option (Figure 6), the signal pulse is sampled continuously by an analogue ring sam-pler such as those used by H.E.S.S. (ARS0) or H.E.S.S. II (SAM) [4]. The camera trigger stops the sampling after a delay, such that pixel signals occurring before and after the trigger are recorded. The readout of the analogue memory is controlled by a field programmable gate array (FPGA). The memory contents are digitized, the peak of the pulse is located, the samples corresponding to ~16 ns around it are summed and this number is read out.

In both options, it is envisaged that two gain channels for each pixel will be used to improve dynamic range. In future simulation studies we plan to investigate the performance of other cell layouts as well as the performance of various arrangements of cells. Optimum trigger conditions for a single telescope are being investigated based on simulation of accidental trigger rates. Furthermore, we are considering Hamamatsu H85000 multi-anode photomultipliers (MAPMT) to reduce per pixel cost. The small size of the MAPMT requires a smaller focal plane scale. Recent work using secondary optics by Vassiliev et al. [17] is encouraging in this respect.

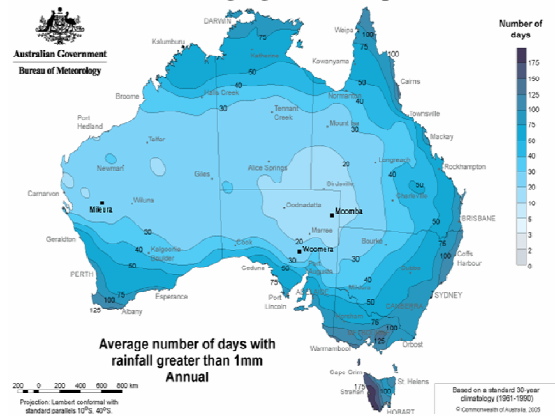

Figure 7: 'rain days' map, www.bom.gov.au/cgi-bin/climate/cgi_bin_scripts/rndays.cgi

Surveying of possible Australian sites is being carried out, the primary consideration is that night sky conditions are dark and cloud free. The best area seems to be near the NE corner of the state of South Australia (Figure 7). Woomera is near this region. The area has low altitudes (< 200m a.s.l.) and is relatively flat with low hills and stony plains in the SW, trending towards sand dunes in the NE. The best sites are likely to be in the south of this region (just to the NE of Marree). The western side has the advantage of being close to the main north-south highway. We will also consider Mileura, in the west of Western Australia, which is one of two sites short-listed for the Square Kilometer Array [16].

## Conclusions

We have carried out simulations of an IACT cell of five telescopes between energies of 1 TeV and 100 TeV. Based on these investigations, we found that large effective collection areas of >1 km$^2$ are achievable using a single IACT cell at multi-TeV energies. Extrapolating this result to an array of 10 independent cells (for example) suggests that TenTen could achieve collecting areas of >10 km$^2$ above 10 TeV, and which could exceed those of H.E.S.S. by a factor of ~50 at ~100 TeV.